\documentclass[%
 reprint,
 amsmath,amssymb,
 aps,
]{revtex4-2}

\usepackage{graphicx}
\usepackage{xcolor}
\usepackage{textcomp}
\usepackage{dcolumn}
\usepackage{bm}
\usepackage{caption}
\usepackage{subcaption}
\usepackage{float}
\usepackage{adjustbox}

\begin{document}

\preprint{APS/123-QED}

\title{Resolving Nonequilibrium Gas Kinetics in Supersonic Neutral Flows with Coherent Rayleigh–Brillouin Scattering}

\author{Atulya Kumar}
\author{Gabriel M. Flores Alfaro}
\affiliation{Advanced Laser Diagnostics and Optical Manipulation Group,
Luxembourg Institute of Science and Technology, 4362 Esch-sur-Alzette, Luxembourg}
\affiliation{Université du Luxembourg, 2 Avenue de l’Université,
4365 Esch-sur-Alzette, Luxembourg}

\author{Marios Kounalakis}
\author{Stefan Karatodorov}
\affiliation{Advanced Laser Diagnostics and Optical Manipulation Group,
Luxembourg Institute of Science and Technology, 4362 Esch-sur-Alzette, Luxembourg}

\author{Alexandros Gerakis}
\email{alexandros.gerakis@list.lu}
\affiliation{Advanced Laser Diagnostics and Optical Manipulation Group,
Luxembourg Institute of Science and Technology, 4362 Esch-sur-Alzette, Luxembourg}
\affiliation{Department of Aerospace Engineering, Texas A\&M University,
College Station, TX 77843, USA}


\date{\today}

\begin{abstract}
We present a characterization of high-speed flows in non-equilibrium thermodynamic conditions using single-shot coherent Rayleigh–Brillouin scattering (CRBS). The technique is applied on a highly under-expanded jet using $\sim200$~ns laser pulses, enabling simultaneous probing of multiple spatial locations. We map the jet’s average axial velocity and density distributions and resolve local velocity gradients, providing access to parameters relevant to turbulence characterization. The measurements are validated against numerical simulations, showing good agreement overall. We find that in most cases individual single-shot spectra exhibit substantial deviation from the bulk-averaged CRBS lineshapes, reflecting the non-Maxwellian velocity distributions generated by shock-induced regimes. The results presented here establish single-shot CRBS as an important tool for direct measurements of flow velocity components, velocity gradients, and density in complex, unsteady supersonic environments.
\end{abstract}

\maketitle


\section{\label{sec:level1}Introduction}

Unraveling the origins of turbulence has remained a central challenge in fluid dynamics for over a century~\cite{reynolds1883xxix,bailly2015turbulence}. By examining the temporal and spatial evolution of key thermodynamic and flow parameters, such as temperature, density, chemical composition, and velocity, researchers can gain deeper insight into the multiscale behavior of fluid and plasma flows~\cite{frisch1990turbulence,fortov2016thermodynamics}. Such understanding directly enables practical advances, including improved control of mixing~\cite{mcdonell2000measurement} and heat transfer in manufacturing processes (e.g., combustion, additive manufacturing, and plasma processing~\cite{merzhanov1996combustion,launder1991current,graves1989plasma}), more accurate prediction of aerodynamic loads~\cite{drela1990method}, drag reduction in ground and air vehicles~\cite{nath2021drag,yoshida2009supersonic}, the optimization of thermal protection~\cite{xie2013thermomechanical}, propulsion efficiency~\cite{ruiz2011vortex}, and structural reliability in spacecraft and high-speed flight systems~\cite{roy2006review}.

To support the development of such technologies, numerous ground-based test facilities~\cite{tunnel9,beresh2015modernization,simms2001nrel} have been established to reproduce a wide range of flow conditions for the evaluation and maturation of systems operating in extreme aerodynamic environments. In recent years, high-speed flows have attracted significant attention in the context of supersonic and hypersonic vehicle development~\cite{bertin2003fifty}. Consequently, advanced optical flow diagnostics are increasingly employed to enable detailed characterization of complex compressible phenomena, including strong shock waves, expansion fans, and highly unsteady mixing processes, which require both high temporal and spatial resolution~\cite{giepman2015high,nel2015schlieren,okhotsimskii1998schlieren}. Underexpanded jets serve as prototypical examples of compressible, high-speed flows characterized by strong shock–shear interactions, expansion waves, and intense turbulent mixing~\cite{FRANQUET201525}. When a gas expands from a high pressure reservoir through a nozzle into a lower pressure environment, the resulting flow field develops a complex system of shock cells, Mach disks, and shear layers that evolve dynamically in both space and time~\cite{FRANQUET201525}. These interactions govern the jet’s acoustic emission, momentum transfer, and mixing efficiency, making underexpanded jets highly relevant to applications such as propulsion systems, rocket exhausts, and supersonic fuel injection~\cite{wu2015three,murugappan2004flowfield,duronio2023under}. Understanding and modeling the extreme dynamics of these flows require detailed, quantitative measurements of instantaneous velocity, temperature, and density fields across a wide range of spatial and temporal scales.

\begin{figure} [h]
\includegraphics[scale =0.31, trim = 2.9cm 7cm 1cm 1cm, clip]{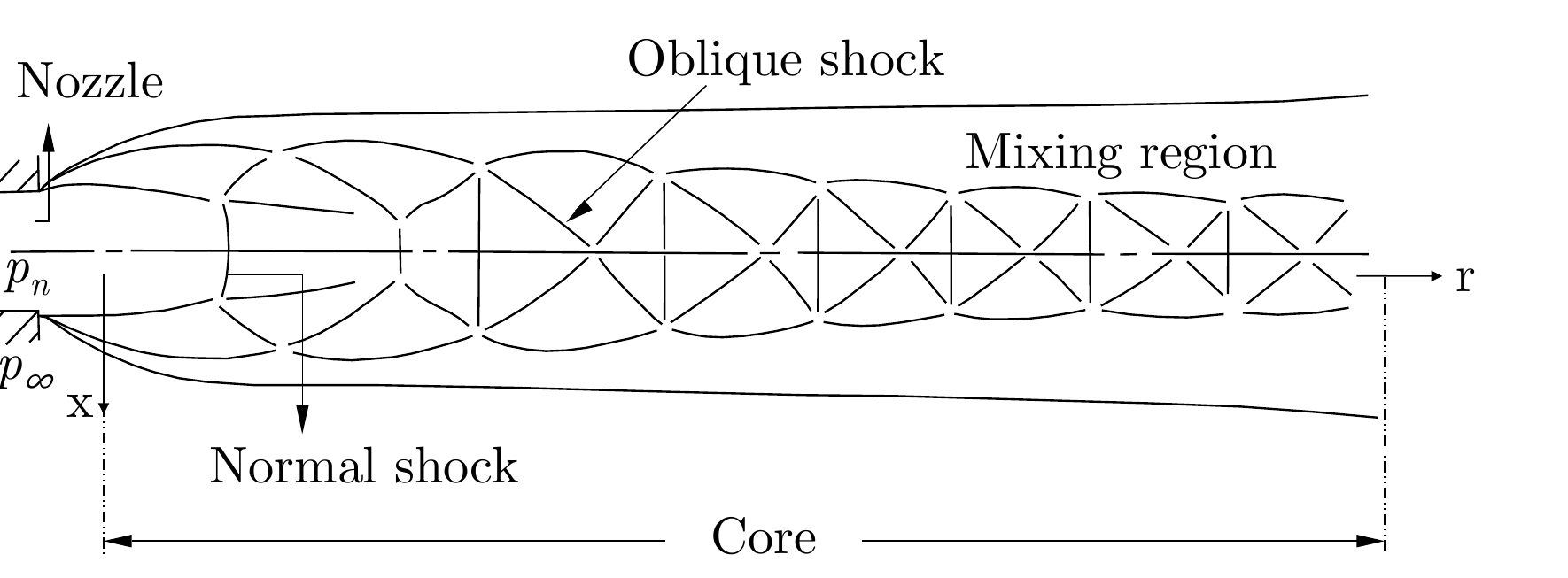}
\caption{\label{fig:jetsch} Schematic representation of the structure of a highly under-expanded jet based on Ref.~\cite{snedeker1964experiments}.}
\end{figure}

In this context, laser-based diagnostics have proven invaluable, providing a largely non-intrusive means of probing flow properties~\cite{danehy2018non} compared to traditional mechanical instruments such as Pitot probes~\cite{chue1975pressure}, hot-wire anemometers~\cite{lomas2011fundamentals}, and thermocouples~\cite{ballantyne1977fine}. While these established tools remain important in certain applications, their use in supersonic and hypersonic flows is often limited by constraints such as intrusive spatial footprints, flow perturbations, and downstream instabilities~\cite{zhang2011measurement}. Importantly, due to their mechanical nature, they can also be damaged by the flow itself. The advent and continuous advancement of laser technology over the past few decades have led to the development of numerous high-power laser systems operating across a broad range of wavelengths and modes, enabling increasingly sophisticated optical diagnostic techniques for high-speed flow characterization~\cite{maiman1960stimulated,steinberg2023optical, miles2015optical, miles2021localized}.

Extensive research has been conducted on underexpanded jets over the past several decades with Particle Image Velocimetry (PIV) techniques~\cite{ANDRE2014188,CaseyTPIV}. Most PIV techniques require seeding with tracer particles that may not accurately follow the gas motion in highly compressible regions~\cite{yuceil2017comparison}. Consequently, there remains a critical need for diagnostic techniques capable of resolving instantaneous, unseeded velocity fields in high-pressure and high–Mach-number flows. 
 
Molecular tagging velocimetry (MTV) alleviates several limitations of particle image velocimetry (PIV) by replacing tracer particles with molecular tracers that are of the same size scale as the fluid medium. MTV enables time-resolved velocity measurements with microsecond-scale temporal resolution~\cite{koochesfahani2007molecular,gendrich1997molecular,bathel2013review}; however, its applicability is constrained by factors such as finite fluorescence lifetimes, species-specific tracers, molecular diffusion, and optical detection challenges associated with the isotropic ($4\pi$ steradian) emission of the tagged species. Krypton tagging velocimetry (KTV) is one such technique that was employed to study underexpanded jets~\cite{Parziale:15}. Another technique called Vibrationally Excited Nitrous Oxide Monitoring (VENOM) was used to characterize similar high speed jet flows~\cite{Hsuvenom}. Femtosecond Laser Electronic Excitation and Tagging (FLEET) represents a more advanced implementation of MTV in which nitrogen molecules, an inherent component of Earth's atmosphere, are electronically excited and subsequently tracked to estimate velocity. By eliminating the need for tracer species, FLEET overcomes many of the chemical and seeding constraints inherent to traditional MTV techniques. 
The application of FLEET for velocimetry was demonstrated in an underexpanded jet~\cite{Michael:11}, highlighting its suitability for high speed, compressible flow regimes. Despite these advantages, FLEET remains subject to many of the same fundamental limitations as established MTV techniques mentioned before.

 To overcome the limitations associated with seeded flow diagnostics, several seedless, laser based techniques have been developed, among which spontaneous Rayleigh scattering is the most prominent. Rayleigh scattering enables non-intrusive measurements of velocity, density, and translational temperature without the need for tracer particles and can be applied to any gaseous medium. One widely adopted implementation is Filtered Rayleigh Scattering (FRS), which infers density and velocity fields by spectrally isolating the Doppler-broadened Rayleigh signal using a molecular absorption filter~\cite{miles2001laser}. 
FRS was used to generate density maps of a co-flowing underexpanded jet~\cite{GeorgeFRS}. Other variants  of Rayleigh scattering based diagnostics have also been applied to underexpanded jets and other canonical flows such as ultraviolet (UV) Rayleigh scattering. UV Rayleigh scattering was used to obtain two-dimensional maps of local gas density across an underexpanded jet using an intensified charge-coupled device (ICCD) camera over a range of stagnation pressures\cite{dam1998imaging}. While Rayleigh scattering techniques provide access to fundamental flow properties—including velocity, density, and translational temperature—they are often limited by weak scattering coefficients leading to low signal-to-noise ratios (SNR), particularly in optically noisy environments. Coherent Rayleigh–Brillouin Scattering (CRBS) which is utilized here, addresses many of these challenges by retaining the measurement capabilities of FRS while offering a significantly enhanced SNR. As a result, CRBS been demonstrated as a promising diagnostic for large-scale experimental facilities, including wind tunnels and plasma environments~\cite{kumar2025multi,randolph2021diagnostics,PhysRevApplied.9.014031}.

 Coherent Rayleigh–Brillouin Scattering (CRBS) is a non-resonant($\chi^{(3)}$) third-order nonlinear optical process that produces a coherently scattered signal whose spectrum encodes the thermodynamic properties of the gas, such as temperature, density, velocity~\cite{Pan_CRBS,SRBS_She,gerakis2016remote}. Owing to its coherent nature, the CRBS signal exhibits minimal beam divergence, enabling efficient signal collection and  greatly improved SNR. Furthermore, its non-resonant nature allows application to a wide range of neutral gases with sufficient polarizability, eliminating the need for tracer species or seed particles. Early implementations of CRBS~\cite{Pan2004_molecular} were unsuitable for investigating transient flow phenomena owing to the long spectral acquisition times (order of $\sim 10-20$ minutes), however recent instrumentation advancements~\cite{Bak:22,Karatodorov2025} have enabled the single-shot CRBS spectral acquisition capability (order of $\sim200$~ns)~\cite{gerakis2013single,gerakis2016remote} and flow velocimetry~\cite{ktvb-mw8z}. 
 
This work demonstrates the advantages of single-shot coherent Rayleigh Brillouin scattering (CRBS) for high-speed, non-Maxwellian flow diagnostics through its unique ability to directly probe the particle velocity distribution within the measurement volume. In this work, the single-shot CRBS technique is employed to measure the axial velocities in different locations of a highly underexpanded gas jet. By implementing a simultaneous, multi-point CRBS detection scheme~\cite{Auk_multiCRBS}, flow velocities at multiple spatial locations are measured simultaneously, enabling the determination of velocity gradients across the flow field within the duration of a single laser pulse.

\section{Experimental methods}

\subsection{\label{sec:level2} Single-shot Coherent Rayleigh Brillouin Scattering}

The single-shot CRBS technique has been extensively described in Ref.~\cite{gerakis2013single, Bak:22, PhysRevA.110.033519}. Only aspects of the operating mechanism relevant to the present experiment are discussed here. In this scheme, two intense laser beams with the same polarization termed the \textit{pumps}, each with intensity $I_{pp_1}$ and $I_{pp_2}$, interfere with each other to produce regions of periodic high and low electric field intensities. Neutral gas particles within the interfering electric field experience a dipole force $F_{D} \propto  \frac{1}{2} \alpha_{eff}\sqrt{I_{pp1}I_{pp2}}$, where $\alpha_{eff}$ is the effective polarizability of the particles of the medium, that forces them to populate the high intensity field regions creating a periodic structure with a density modulation termed an \textit{optical lattice}. The optical lattice period is given by, 
\begin{equation}
    \lambda_g = \frac{\lambda_{pp}}{2sin(\phi/2)},
\end{equation}
where $\lambda_{pp}$ is the wavelength of the pumps and $\phi$ is the crossing angle.
A third beam, called the probe $I_{p}$, is introduced to interact with the  optical lattice at the Bragg condition, $\lambda_{p}=2\lambda_gsin(\theta_b)$ where $\theta_b$ is the Bragg angle, producing a coherently scattered signal beam~$I_{signal}$. This signal beam carries spectroscopic information related to the thermodynamic properties of the particles within the lattice.

\begin{figure}[ht]
\includegraphics[scale =0.45, trim = 6cm 4cm 11cm 3cm, clip]{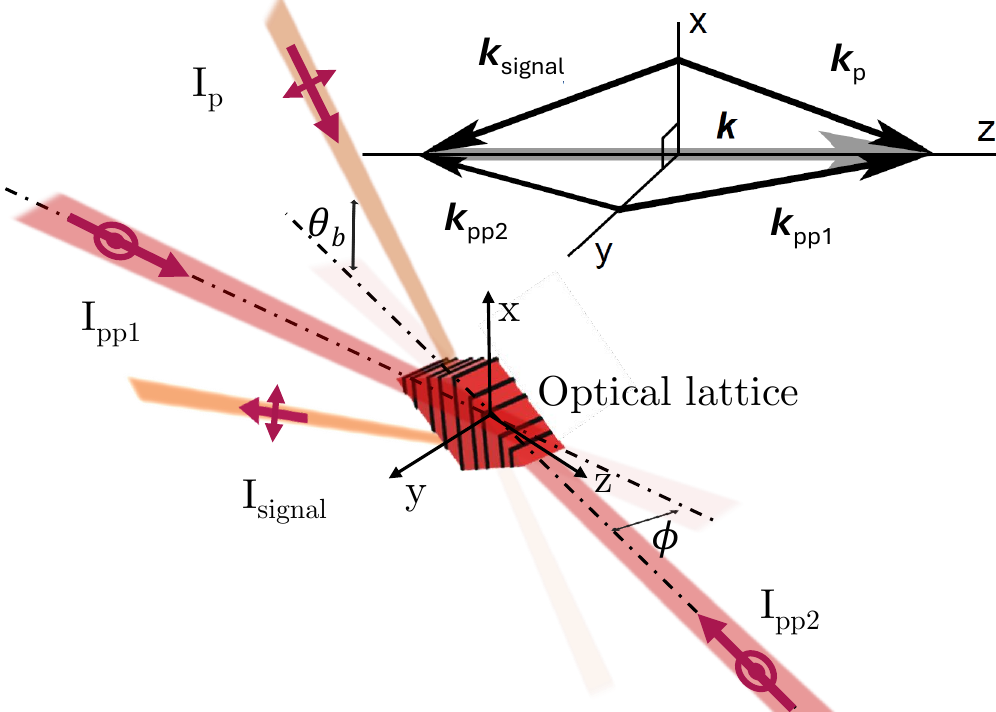}
\caption{\label{fig:lattice} The four wave mixing process ($\chi^{(3)}$), illustrating the interaction of the pump $I_{pp_1}$, chirped pump $I_{pp_2}$, and probe beams $I_{p}$ that generates the CRBS signal $I_{signal}$.}
\end{figure}

 If the pump beams have a relative frequency difference $\Delta f$, the lattice would interact with particles with  phase velocity given by the equation,
 \begin{equation}
    v_{ph} = \frac{\lambda_{pp}\Delta f}{2 sin(\phi/2)}.
    \label{eqn:vphase}
\end{equation}
This corresponds to a subset of particles of $v_{ph}$ manifold in the VDF of the lattice volume. By varying the frequency in time, i.e. by introducing a chirp $\Delta f(t)$, one generates a time-dependent phase velocity $v_{ph} (t)$ to be swept across in velocity space, effectively scanning the full VDF of the lattice volume. In practice, this is achieved by chirping one pump pulse relative to the other, such that the entire scan is completed within the duration of a single laser pulse of about $\sim 200$ns.

\begin{figure*}[hbt!]
\includegraphics[scale =0.5]{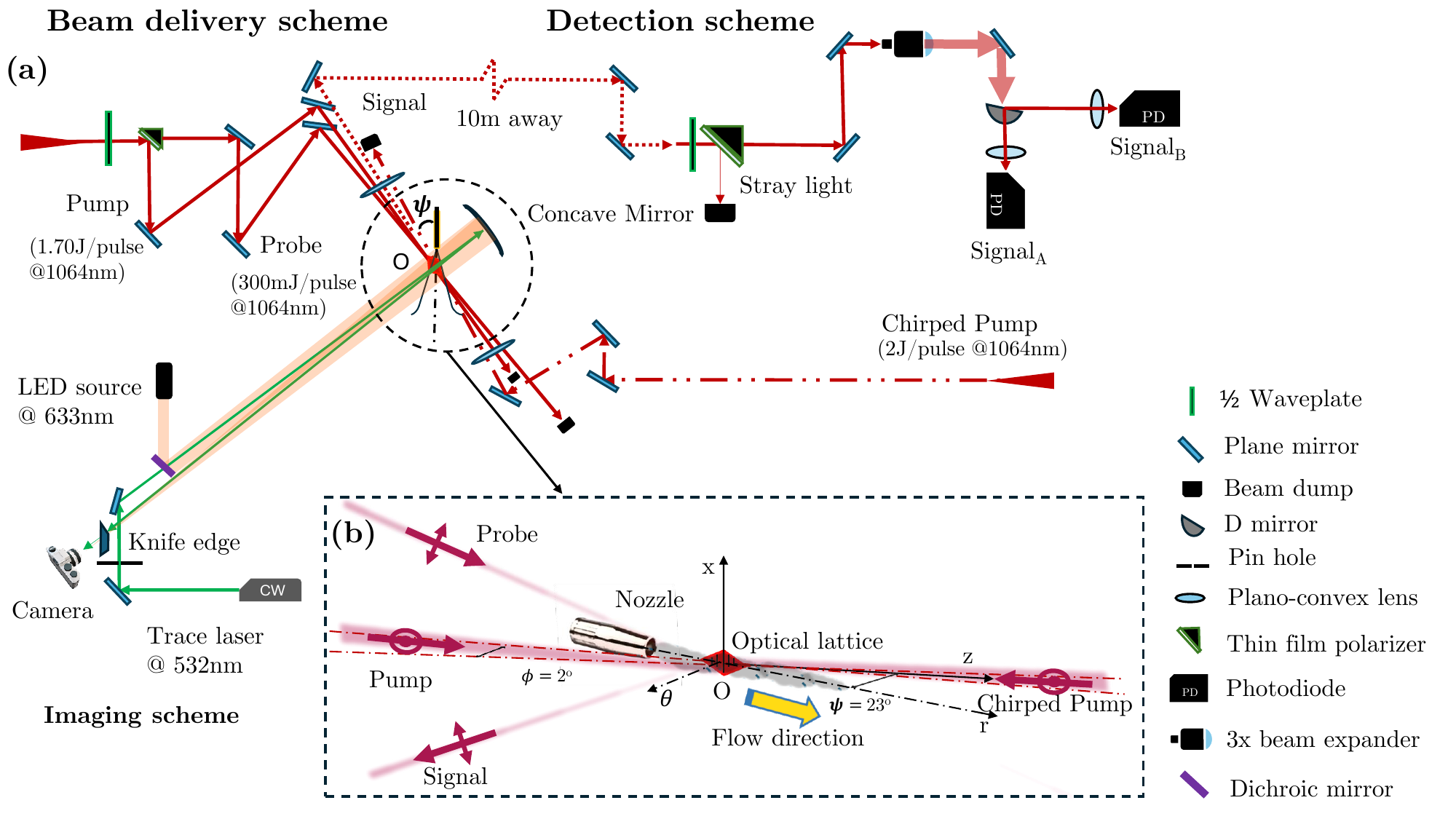}
\caption{\label{fig:beam} (a) Optical setup of the experiment. (b) Orientation of the jet with respect to the laser beams.}
\end{figure*}

\subsection{\label{sec:level2} Experimental Apparatus}

The laser system employed in this manuscript for single-shot CRBS is a custom-built setup, with detailed specifications available in Refs.~\cite{Bak:22,Karatodorov2025}. Here, only the relevant output characteristics are summarized. The pumps, and probe beams deliver pulse energies of $\sim2.0$~J, and $0.3$~J, respectively, at a wavelength of $1064$~nm. The chirp rate applied to the pump beams is approximately $13$~MHz/ns, and all laser pulses have a temporal duration of about $230$~ns. The three beams are directed through a set of $500$~mm lenses, which focus them into the probe volume in a counter-propagating folded BOXCARS configuration, as illustrated in Fig.~\ref{fig:lattice}. The Rayleigh range of the beams was experimentally measured to be $\sim30$~mm. With the help of a beam profiler (Ophir BGP-G-SP504S), the beam overlap distance was measured to be $1~$mm and the probe diameter is $100~\mu$m. The underexpanded jet that is being characterized located at the focal plane and is generated by a nozzle oriented at an angle of $\psi = 23^o$ with respect to the optical axis of the lenses so as to not obstruct the interacting laser beams. The nozzle has a diameter of $\sim3.6$~mm and is connected to a reservoir of air regulated to provide a nozzle pressure ratio ($NPR=p_n/p_\infty$) of $\sim6.5$ where $p_n$ is the reservoir pressure and $p_\infty$ is the ambient pressure. 

Fig.~\ref{fig:beam}(a) shows a schematic of the beam delivery, detection  and imaging schemes employed here. The beams coming from the laser system follow an experimental geometry as indicated in the beam delivery scheme. The CRBS signal beam emanating from the probe volume is routed to the signal detection setup located $\sim10$~m away. In the signal detection setup, the signal beam is first passed through a TFP-$\lambda/2$ waveplate and then through a $3\times$ beam expander. This expansion increases the beam diameter, improving the spatial resolution and overall resolving capability of the detection system. After the expander, the beam is divided into two parts using a 'D'-shaped pick-off mirror. The D-mirror is oriented horizontally, thereby dividing the signal beam into two 'beamlets'. Each corresponds to a distinct probe volume along the x direction of the interrogation region (see Fig.\ref{fig:beam}(b)). Each 'beamlet' is then directed and focused into two independent fast InGaAs photodiodes (Hamamatsu G6854-01) referred to as $PD_A$ and $PD_B$. We refer to Ref.~\citenum{Auk_multiCRBS} for more information and details about the 1D CRBS detection scheme employed here.

\begin{figure} [h]
\centering
\includegraphics[scale =0.42, trim = 6cm 0cm 1cm 0cm, clip]{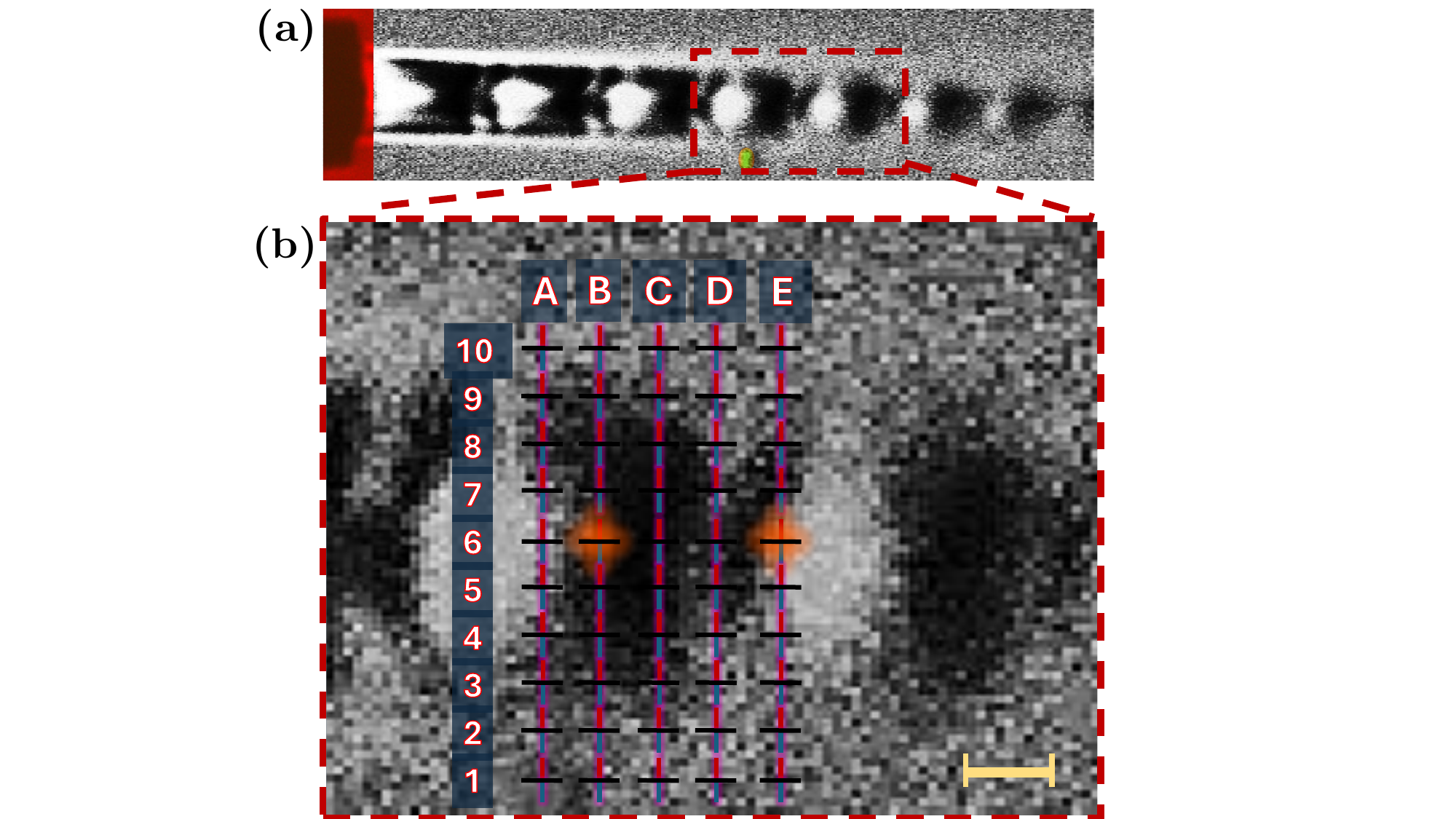}
\caption{\label{fig:jet}(a) A schlieren image of the under-expanded jet with $6.5$ NPR. The box indicates the region of interest of the jet and the green circle highlights the tracer. (b) All probe locations are marked with respect to  the schlieren image. Blue and red lines indicate locations corresponding to $PD_A$ and $PD_B$ respectively. Location B6 and E6 have been highlighted for further discussions. The yellow scale bar indicates $1$~mm of length.}
\end{figure}

A double-pass schlieren imaging scheme is additionally implemented to observe the locations while being probed with CRBS~\cite{settles2001schlieren}. As shown in Fig.~\ref{fig:beam}(a), a $633$~nm LED illumination is directed to a concave mirror ($f_l=500$~mm), located just behind the interrogation region, with the help of a 'red hot' dichroic mirror (Thorlabs FM02). The concave mirror focuses the light from the LED on a knife edge and then relays into a charged couple device (CMOS) camera sensor (Basler acA3088). A continuous wave laser at $532$~nm wavelength, referred to as 'tracer', is directed to the probe volume through the dichroic mirror and focused on to the corresponding location on the camera sensor. The schlieren image of the under-expanded jet the tracer is shown in Fig.\ref{fig:jet}(b).

The experimental campaign consisted of ten vertical scans across the jet at five equally spaced axial locations (refer Fig.~\ref{fig:jet}(b)), covering a region that spans an entire shock cell. At each location, a total of $20$ single-shot CRBS spectra were recorded at a repetition rate of $5$~Hz. 

\section{Data processing and Results}

It is important to note that photodiodes $PD_A$ and $PD_B$ record spectra from two vertically adjacent locations ($x$-direction in the lattice frame and $r$-direction in the nozzle frame) within the interrogation region~\cite{Auk_multiCRBS}. The probe volume of each beamlet has a diameter of approximately $50~\mu$m, and its spectrum reflects the averaged velocity distribution function (VDF) of particles within it. Using this information, all acquired spectra from all beamlets can ultimately be mapped simultaneously in the velocity space to possibly reconstruct the VDFs of all particles across the total probe volume.

Single-shot CRBS spectral acquisition is performed in the time domain. Since the spectra are recorded by photodiodes and processed on an oscilloscope, they must be transformed into velocity space for meaningful interpretation. This transformation is carried out by determining the instantaneous frequency of the chirped laser pulse over its temporal duration through a heterodyne detection system between the two pump beams. For each spectrum, the corresponding heterodyne signal is analyzed using a fast Fourier transform (FFT) algorithm to convert the time domain signal to the frequency domain, followed by a linear fit to extract the instantaneous frequency $\Delta f (t)$, across the pulse duration. The mean value of $\Delta f (t)$ from all heterodyne acquisitions throughout the experimental campaign is then used to map each spectrum from frequency domain into the velocity domain through Eqn.~\ref{eqn:vphase}. The flow velocity within the probe volume is measured using the following expression based on the kinetic theory of gases~\cite{LIFSHITZ19811},
\begin{equation}
V = \frac{{\displaystyle \int v_{ph} f(v)\, dv}}{{\displaystyle \int f(v)\, dv}}
\end{equation}
where $ f(v)$ is the spectral profile and $v_{ph}$ is the phase velocity. Since the nozzle is inclined by an angle $\psi$ with respect to the optical axis of the lattice, the measured flow velocity reflects the projection of the local velocity vector field within the probe volume onto this axis. In applying the angular correction, it is assumed that the radial component of the velocity field is small relative to the axial component and can therefore be neglected. Under this assumption, the axial flow velocity is obtained by $v_{z} = V/\cos\psi$. For the calibration spectra in quiescent conditions, the magnitude of the flow velocity yields $0$~m/s as expected. For more information on single-shot CRBS velocimetry the reader is referred to Ref.~\citenum{ktvb-mw8z}.

\begin{figure} [hb!]
\includegraphics[scale =0.47, trim = 0cm 0cm 0cm 0cm, clip]{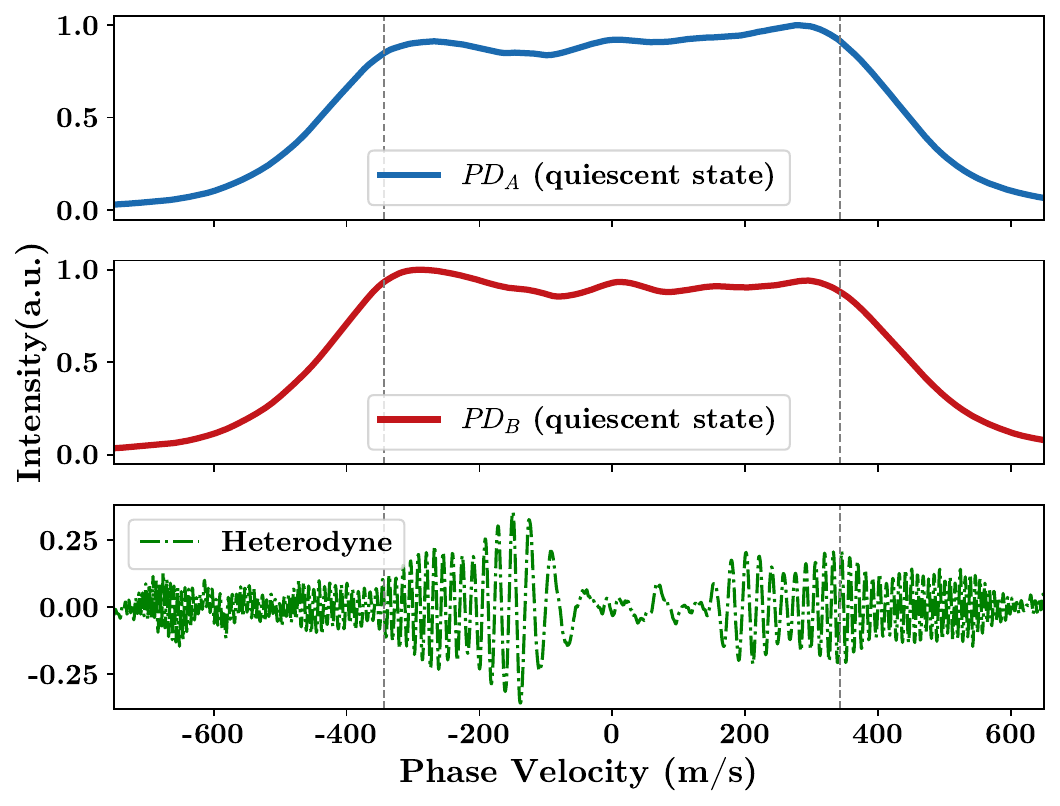}
\caption{\label{fig:cal_spec} Single-shot CRBS spectra from photodiodes $PD_{A}$ and $PD_{B}$ in quiescent conditions. The dotted gray lines represent the local speed of sound in air. The dotted green curve shows the mean heterodyne used for velocity mapping.}
\end{figure}

\begin{figure}[ht!]
    \centering
    \includegraphics[width=0.49\textwidth, trim = 0cm 0cm 0cm 0cm, clip]{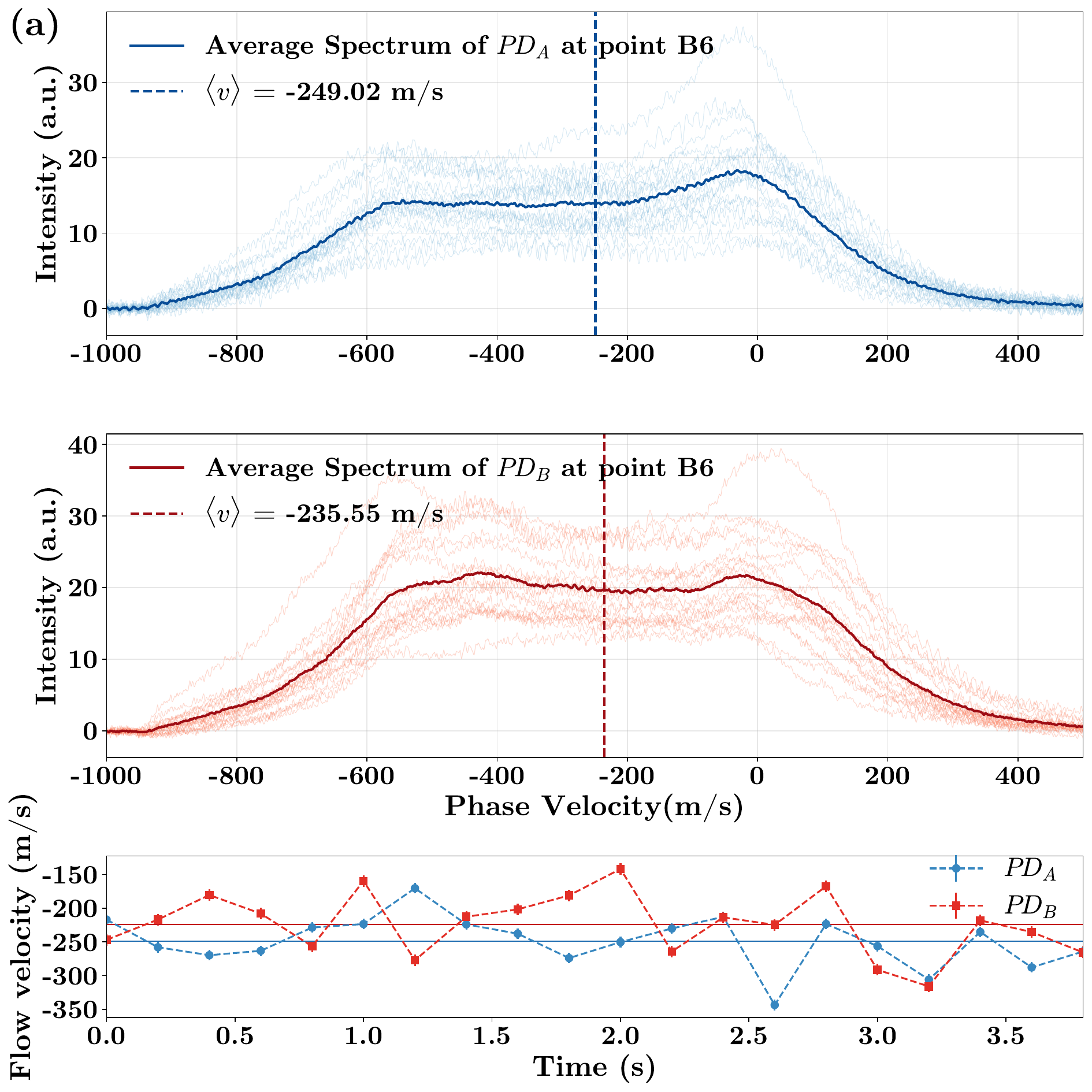}\\[0.5em]
    \includegraphics[width=0.49\textwidth, trim = 0cm 0cm 0cm 0cm, clip]{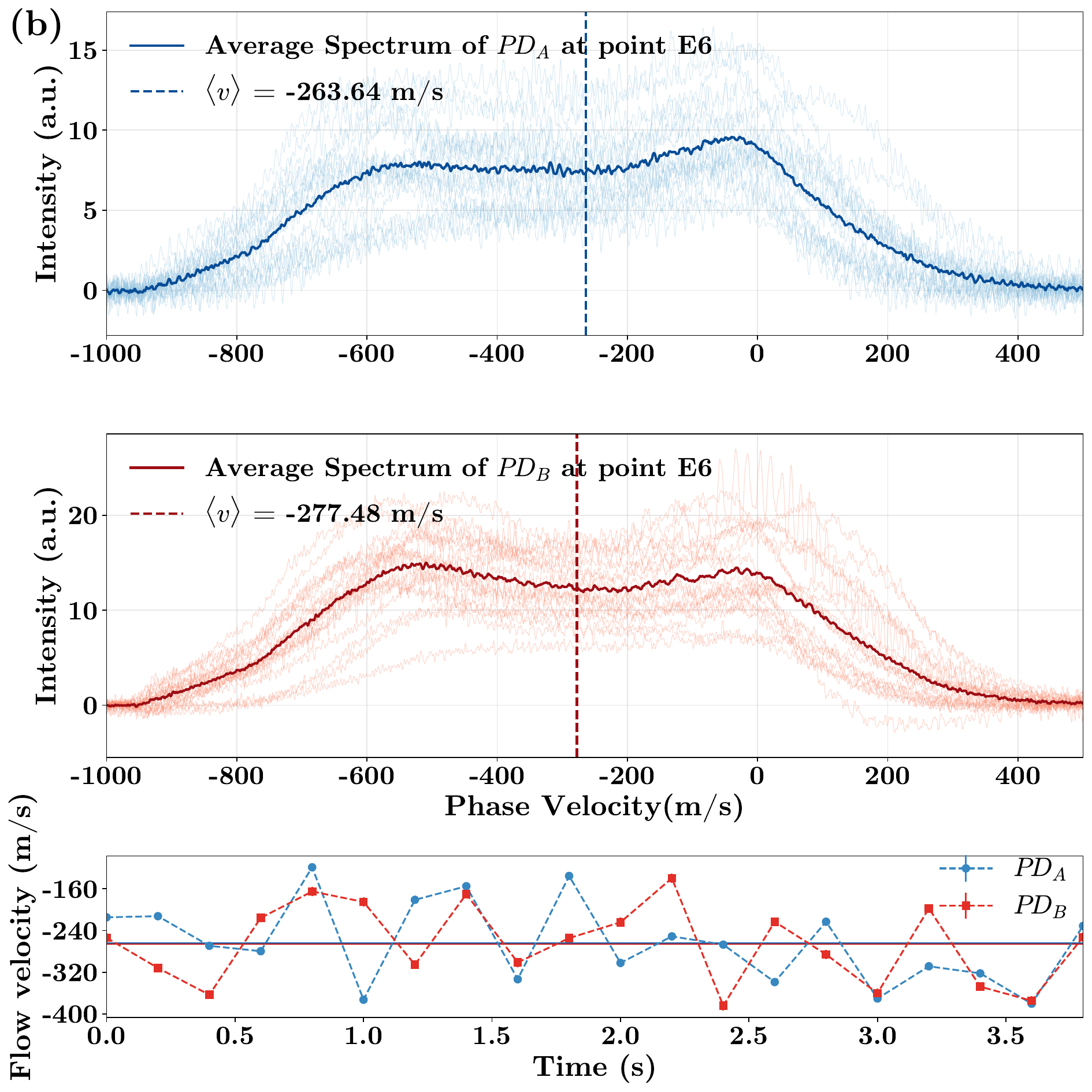}
    \caption{The spectral spread of points B6 (a)  and E6 (b) (refer Fig.\ref{fig:jet}(b)). The darker curve represents the CRBS spectrum averaged over $20$ laser pulses. The scatter plots below show the variation in velocity at each point for every acquisition at a repetition rate of $5$ Hz. The horizontal solid lines indicate the mean velocity across all acquisitions for each photodiode. For a single-shot measurement, the error bars are within $\pm5$ m/s for all acquisitions.}
    \label{fig:ave_spec}
\end{figure}

For this experiment, an underexpanded jet of NPR of $6.5$ was selected to ensure the presence of strong, well-defined density gradients sufficiently downstream of the nozzle exit. This allows the prominent flow features to lie within the focal plane while keeping the nozzle itself out of the path of the lasers, thereby avoiding unwanted scattering of the beams. Fig.~\ref{fig:jet}(b) indicates the probe locations within cells $4$ and $5$ of the underexpanded jet. The red and blue markers indicate the probe locations corresponding to $PD_A$ and $PD_B$, while the markers highlighted in orange denote the positions whose spectra are examined in more detail in the following sections. Each marker represents $20$ single-shot CRBS spectral acquisitions obtained from that specific location in the jet. Figures \ref{fig:ave_spec}(a) and \ref{fig:ave_spec}(b) show the shot-to-shot spectral variations and compare them with the 20-shot averaged spectra for both photodiodes. These comparisons reveal the degree of instantaneous velocity fluctuation present within the underexpanded jet.
The direction of the spectral shift is important, as it reflects the true average direction of motion of particles within the flow. Accordingly, a sign convention must be defined based on the experimental configuration and a known reference flow direction. This is particularly important when considering spectra acquired near the mixing regions of the jet as velocity shifts occur opposite to the primary flow direction, indicating localized backward motion consistent with flow recirculation.

\begin{figure}[hbt!]
 \includegraphics[width=0.42\textwidth]{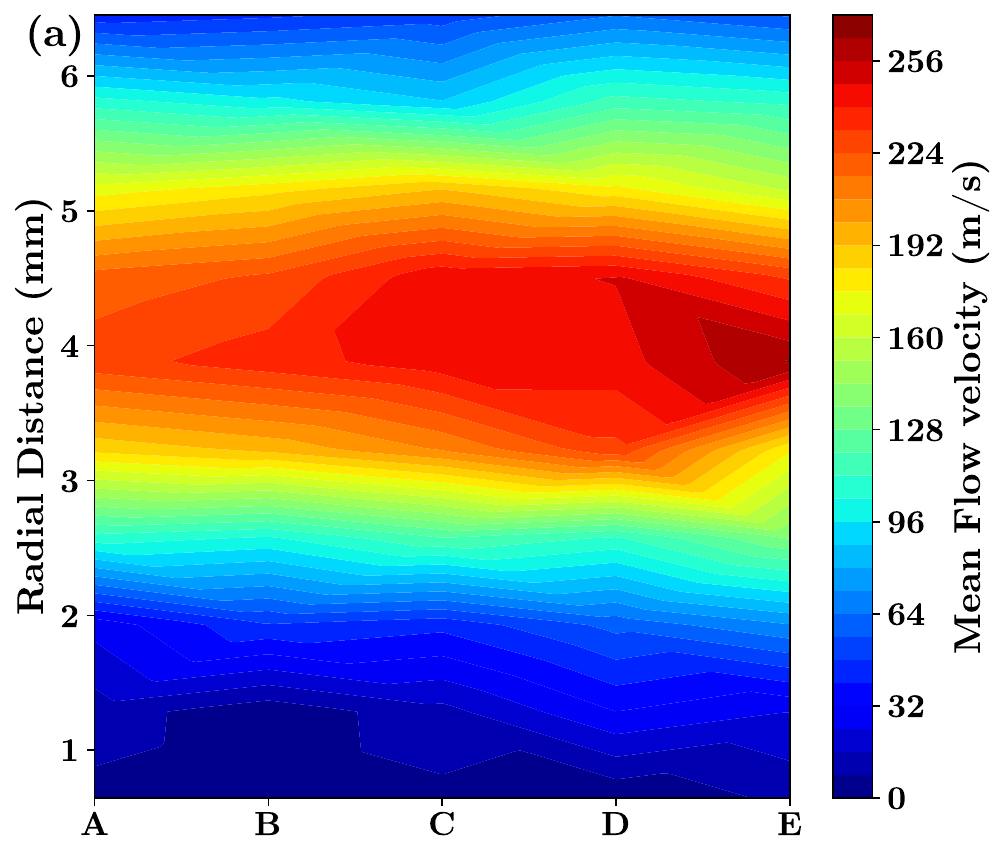}\\[1em]
    \includegraphics[width=0.42\textwidth]{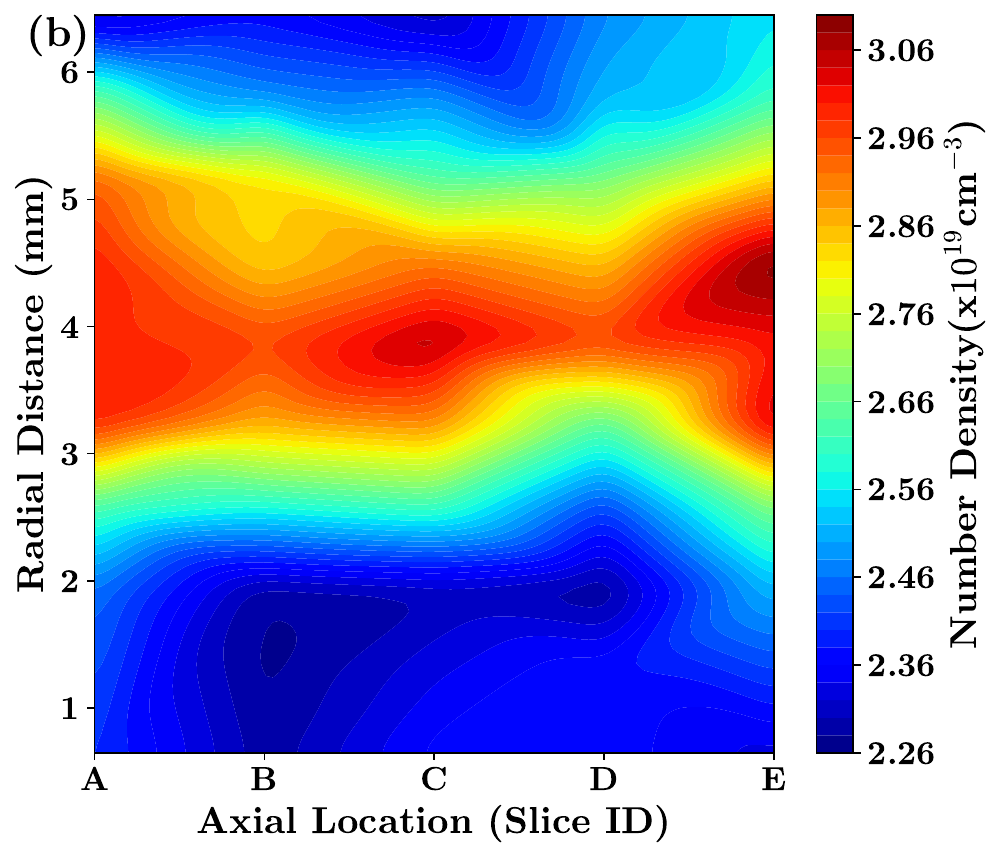}
\caption{\label{fig:cont}(a) Average velocity map of the probed locations in the jet obtained using CRBS.
(b) Average density map across the jet derived from the CRBS measurements.}
\end{figure}

Based on the two-point measurements, a velocity map is constructed and presented in Fig.~\ref{fig:cont}. Although two-point detection is not strictly required— since single-point detection could achieve the same result at twice the acquisition time— the true advantage of the two-point CRBS approach is discussed in the subsequent sections. Fig.~\ref{fig:cont}(a) presents the average axial velocity map, derived from the mean spectra recorded by the two photodiodes at each probed location of the jet. Each photodiode measurement area corresponds to a probe volume of $50~\mu m$, and the resulting average velocities have been smoothed using Gaussian interpolation between measurement points. Additionally, Fig.~\ref{fig:cont}(b) displays the corresponding particle number densities calculated from the same spectra using the relation described below,
\begin{equation}
n = n_{amb}\sqrt{\frac{\displaystyle \int f(v)\, dv}{\displaystyle \int f_{amb}(v)\, dv}}
\end{equation}
where $n_{amb}$ is the ambient number density and $f_{amb}(v)$ is the reference spectral profile that is experimentally obtained prior to performing the measurement. It should be noted that the density estimation is sensitive to the relative intensity of the spectrally integrated reference spectrum. A comparison of Fig.~\ref{fig:cont}(b) with the schlieren image in Fig.~\ref{fig:jet}(a) shows that the reconstructed, averaged density contours closely reproduce the key flow features, particularly in slices C, D, and E (see Fig.~\ref{fig:jet}(b)). The averaged velocity contours in Fig.~\ref{fig:cont}(a) further corroborate this correspondence, exhibiting relatable behavior in the same regions.

Furthermore, as the jet is scanned across the CRBS probing volume, it passes through high-density regions that generate strong refractive index gradients, leading to beam-steering effects. This steering introduces additional optical background to the signal. To correct for this effect, the background level at each probe location is measured by blocking the probe beam and recording the remaining background traces. The measured background is then appropriately scaled and subtracted to remove the pedestal underlying the signal thus isolating the true spectral response. For further details on the background subtraction procedure, the reader is referred to Ref.~\citenum{Auk_multiCRBS}.

\section{Discussions}

\begin{figure}[hbt!]
\centering
    \includegraphics[width=0.45\textwidth,trim = 0cm 1cm 0cm 0cm, clip]{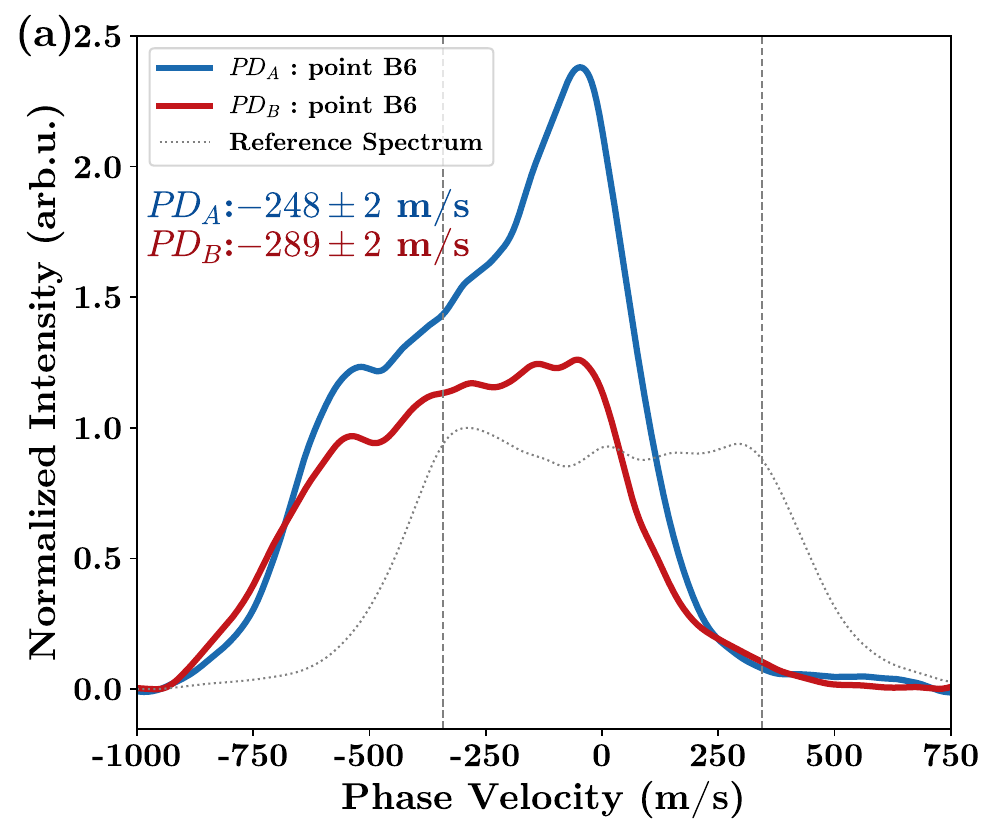}\\[0.1em]
    \includegraphics[width=0.45\textwidth]{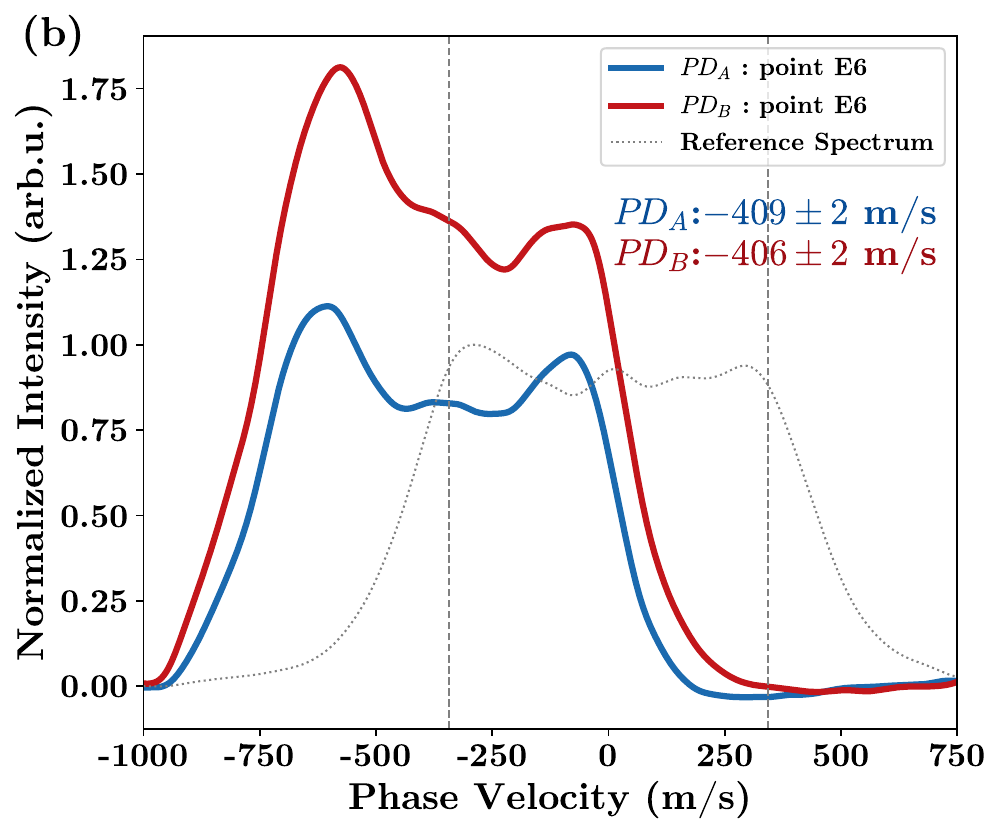}
\caption{\label{fig:SSspec}Single-shot CRBS spectra from points B6 (a) and E6 (b), located within the core of the underexpanded jet, shown alongside the corresponding reference spectrum. The two spectra at each location were recorded simultaneously. A second-order Butterworth low-pass filter was applied to enhance clarity and normalized with the average reference spectrum.}
\end{figure}

From the collection of single-shot CRBS spectra shown in Fig.~\ref{fig:ave_spec}, it is evident that the averaged spectra obtained from the $20$ individual measurements provide a poor representation of the jet dynamics. Since the under-expanded jet exhibits numerous unsteady flow features and strong velocity gradients, single-shot CRBS spectral acquisitions are essential for accurately capturing these characteristics. This also demonstrates the importance of single-shot CRBS spectral acquisition. Fig.~\ref{fig:SSspec} presents selected single-shot spectra from the same locations as the averaged spectra. Many of these spectra reflect non-Maxwellian velocity distribution functions, particularly in regions \textit{B6}(Fig.~\ref{fig:SSspec}(a)) and \textit{E6}(Fig.~\ref{fig:SSspec}(b)). A closer comparison with the schlieren images (shown in Fig.~\ref{fig:jet}b)) indicates that these locations correspond to regions probed in the vicinity of shock waves with \textit{B6} on the aft and \textit{E6} on the fore.

Consider the spectra in Fig.~\ref{fig:SSspec}(a) as an example. The CRBS spectrum from $PD_A$ exhibits a prominent right Brillouin peak centered near $-60$~m/s; although the mean particle velocity at this location is about $\sim-250$~m/s, a substantial fraction of the population is concentrated around $\sim-60$~m/s. The high particle number density and the presence of this subsonic population within the VDF suggests that the probed region lies in the vicinity of a strong shockwave. Similar features are observed in other spectra obtained from the jet core. As the probe volume shifts toward the shear layers, this behavior diminishes, and signatures of recirculation — seen as oppositely directed spectral shifts between the two photodiodes—emerge, consistent with expectations as previously reported in Ref.~\cite{Auk_multiCRBS}. 

In contrast, from Fig.\ref{fig:SSspec}(b), the spectrum from $PD_B$ corresponds to a region of higher density compared to that observed by $PD_A$. The dominance of the left Brillouin peak over the other spectral features indicates that the associated probe volume lies in the vicinity of a shockwave, consistent with the schlieren observations. Meanwhile, the spectrum from $PD_A$ appears more symmetric about the Rayleigh peak, suggesting that it probes a region exhibiting isentropic supersonic flow with an estimated Mach number of approximately $1.2$. 

This is possibly due to the parameters that govern the CRBS spectrum, mainly the gas-dynamic conditions within the probe volume and the dimensionless y-parameter~\cite{Pan2004_molecular}. The y-parameter relation is as follows:
\begin{equation}
y \propto\frac{\lambda_g}{\Lambda}
\end{equation}
where $\lambda_g$ is the optical lattice wavelength and $\Lambda$ is the mean free path of the particles in the vicinity of the lattice. With the $\lambda_g$ fixed, the shock wave imposes a steep density gradient over a length scale orders of magnitude smaller than the probe volume. As the density increases, the particle mean free path $\Lambda$ decreases accordingly. These rapid spatial variations in density and thus in $\Lambda$, cause the local y-parameter to vary across the region being probed. As a result, the measured CRBS spectrum becomes a convolution of lineshapes associated with different y-parameter regimes over the probe volume of these shock affected flows.

In order to assess the single-shot CRBS measurements, a simulation of the underexpanded jet was performed. Although conventional mechanical diagnostics such as flow meters or Pitot probes could, in principle, be used to measure velocities, their use in this flow environment would be problematic: introducing these instruments into the jet generates strong shock waves upstream of the probe tip, significantly altering the local flow and compromising the accuracy of the measurement. While such disturbances can be modeled theoretically, the probe volumes in this study are on the order of hundreds of microns — far smaller than the spatial footprint of mechanical probes — rendering them unsuitable for meaningful comparison. For these reasons, simulation-based validation was chosen as the more reliable approach.

The simulations were carried out in OpenFOAM using the Reynolds-averaged Navier–Stokes (RANS) solver \textit{rhoCentralFoam}~\cite{openfoam}. Although Large Eddy Simulation (LES) generally provides more accurate and physically detailed turbulence modeling for underexpanded jet flows, the RANS approach was chosen here because it is significantly less computationally demanding, readily available within OpenFOAM, and has been demonstrated in previous studies of underexpanded jets~\cite{zang2017openfoam}. An axisymmetric model of the underexpanded jet was constructed using the same boundary conditions as the experimental nozzle. Each discrete element of the mesh defining the domain was approximately $150\mu$m in size, providing relatively similar spatial resolution as that the measurements. The simulation was run for a total flow duration of $20$ms in time steps of $50\mu$s.

\begin{figure}[hbt!]
\centering
     \includegraphics[width=0.48\textwidth,trim = 0cm 13cm 0cm 0cm, clip]{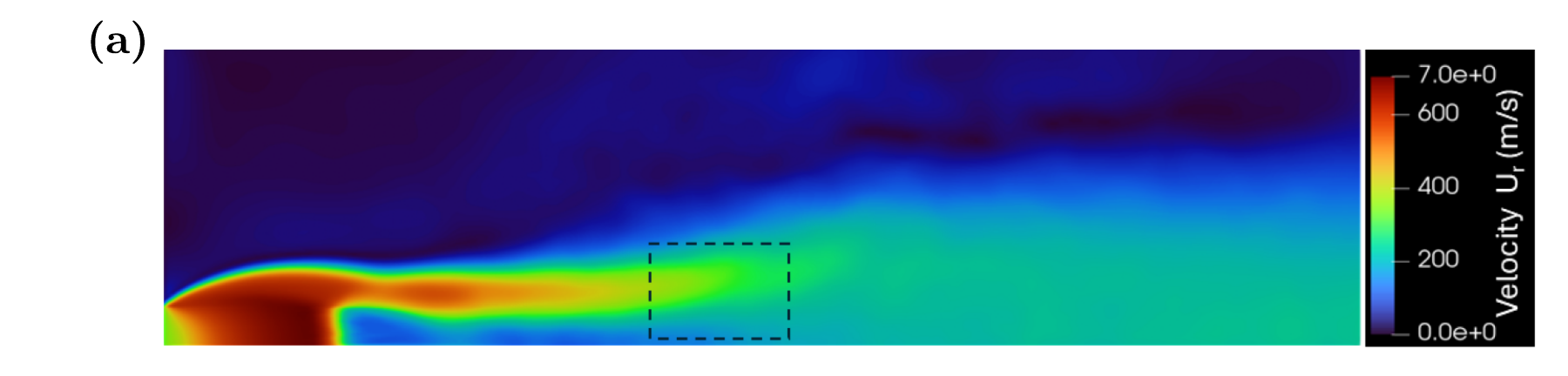}\\[0.1em]
    \includegraphics[width=0.48\textwidth,trim = 0cm 0cm 0cm 8cm, clip]{simvsreal.pdf}
\caption{\label{fig:simvreal}(a) Time averaged velocity contour plot of the under expanded jet. The black box indicates the region around the $4^{th}$ and $5^{th}$ shock cell of the jet. (b) Comparison of the simulation results with the mean CRBS measurements across the jet. The error bars indicate $\pm1$ standard deviation from the average value of each measurement.}
\end{figure}

Fig.~\ref{fig:simvreal} shows the comparison of the averaged measurements with time averaged simulation results of the underexpanded jet. Determining the exact measurement locations in the simulation relative to the experiment can be challenging, as there were no distinct physical flow features to serve as reference points. As a result, the starting location was roughly estimated from the beginning of the fourth shock cell of the jet, with an uncertainty of approximately $\pm0.5$~mm, corresponding to the width of the tracer. Fig.~\ref{fig:simvreal}(b), shows the variation of averaged axial velocity in the jet for corresponding scan lines shown in Fig.\ref{fig:jet}(b) for both photodiodes. The measured average velocities generally fall within the range of the simulated averages, with the exception of slice $9$, located in the mixing region. Although most measured velocities agree with the simulated values on average, a slice-by-slice comparison reveals that the experimental measurements tend to be slightly higher than their simulated counterparts. This discrepancy is likely due to the RANS simulation’s tendency to over-dissipate the flow and the limitations of the turbulence model employed especially in regions away from the core of the flow\cite{bensow2006comparative}. 

\begin{figure}[hbt!]
\centering
    \includegraphics[width=0.5\textwidth,trim = 0cm 0cm 0cm 0cm, clip]{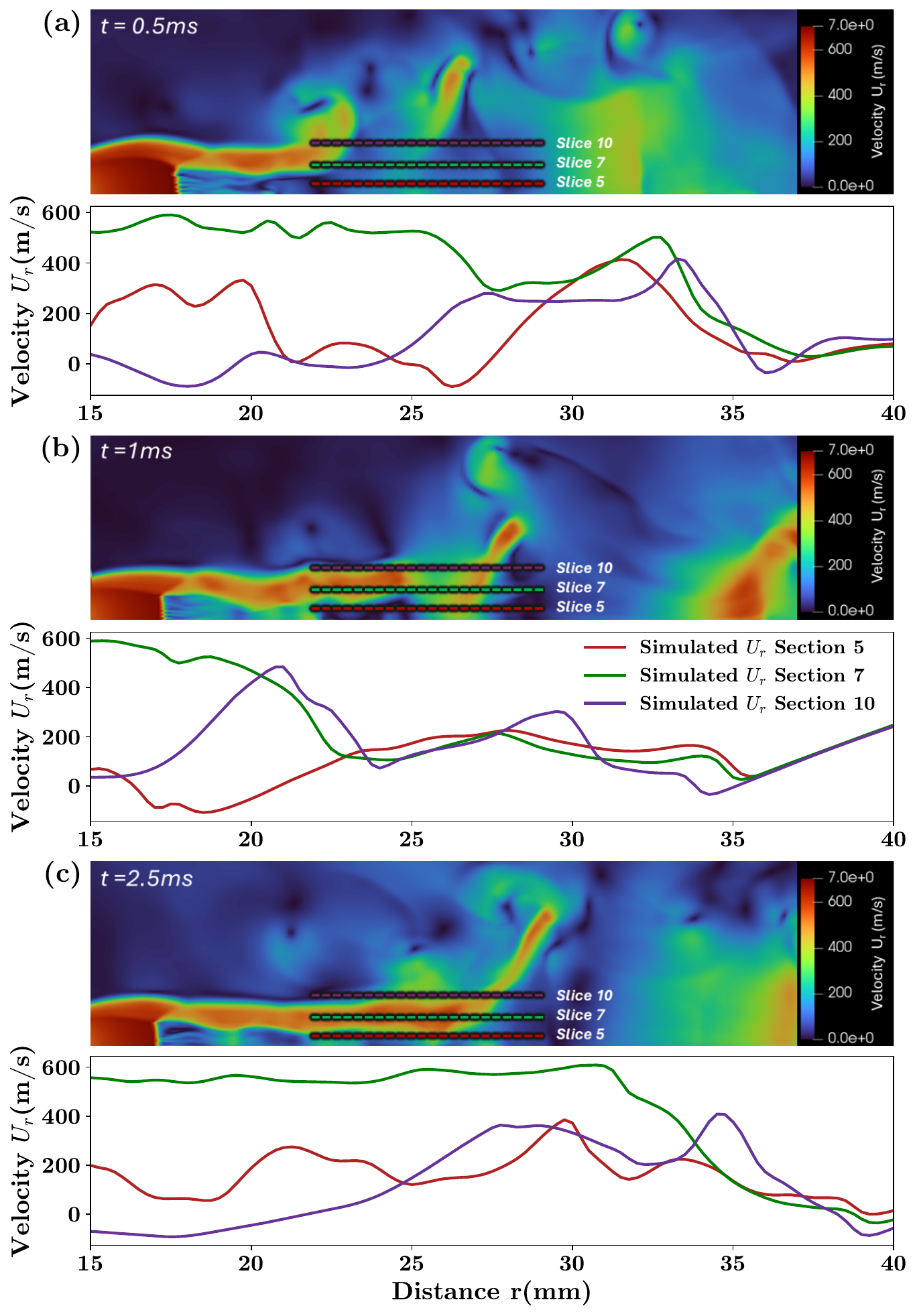}
\caption{\label{fig:V_fluc}Simulated velocity fluctuations over time along each slice corresponding to the probe volume.(a), (b), and (c) show the evolution at time intervals of $500 \mu$s.}
\end{figure}

A direct comparison of single-shot velocities is not appropriate because the experimental measurements were acquired at a $5$Hz repetition rate, while the simulations were performed with a time step of $50\mu$s (or $20$kHz). Figure~\ref{fig:V_fluc} shows the axial velocity variations for three slices over a period of up to $2.5$ms. The plots indicate that the simulated axial velocity fluctuations closely resemble those observed in the experimental single-shot measurements. 

Temperature measurements were not pursued in this work since many of the recorded spectra are distinct reflections of non-Maxwellian VDFs and deviate substantially from standard single-shot CRBS lineshapes observed in quiescent conditions~\cite{gerakis2013single}. Under such conditions, extracting a reliable translational temperature becomes highly uncertain and does not yield physically meaningful values under the Maxwellian definition of temperature~\cite{landau1981statistical}. Nevertheless, variations in spectral width are still observable across the dataset. In particular, spectra associated with post-shock regions exhibit noticeably broader features, qualitatively indicating a phenomenological increase in translational temperature. A rigorous quantification of temperature in these non-equilibrium conditions, however, requires a more general definition for temperature, a sophisticated analysis framework and is therefore beyond the scope of the present study.

One of the main motivations for implementing two-point detection, in addition to acquiring single-shot spectra, is to quantify parameters that depend on velocity gradients—instantaneous vorticity being a key example. This measurement represents an important step toward characterizing turbulence in such unsteady flows using single-shot CRBS. In cylindrical coordinates, and under the assumption of axisymmetry, the vorticity vector reduces to a single non-zero azimuthal component,
\begin{equation}
    \omega_\theta = \frac{\partial v_r}{\partial z} - \frac{\partial v_z}{\partial r}.
\end{equation}

Since the single-shot measurements are acquired at two adjacent locations simultaneously along the $r$-direction, with characteristic length of $50~\mu$m, a lower bound estimate of the vorticity in the $\theta$-direction relative to the jet can be estimated. In this analysis, the assumption $v_r$ $<<$ $v_z$ is made based on supporting simulation results. Table.~\ref{tab:table4} presents the average estimated vorticity component, $\omega_{\theta}$ of the jet from the core (Slice 6) and the shear layer (Slice 10) of the jet.

\begin{table}[hbt!]
\caption{\label{tab:table4}%
Comparison of average $\theta$-direction vorticity and its standard deviation.}
\centering

\resizebox{\columnwidth}{!}{%
\begin{tabular}{lccccc}
\hline\hline
$\omega_\theta\times10^{5}\ (\mathrm{s}^{-1})$ & A & B & C & D & E \\
\hline
Slice 6:  & $-7.5\pm15.3$  & $-5.6\pm11.6$  & $-2.8\pm20.5$  & $-7.3\pm23.3$  & $-1.3\pm20.4$ \\
Slice 10: & $7.5\pm10.7$ & $5.3\pm7.3$ & $6.5\pm9.2$ & $-4.4\pm9.0$  & $1.4\pm8.5$ \\
\hline\hline
\end{tabular}
}

\end{table}

\begin{figure}[hbt!]
\centering
    \includegraphics[width=0.5\textwidth,trim = 4cm 0cm 2.5cm 12.72cm, clip]{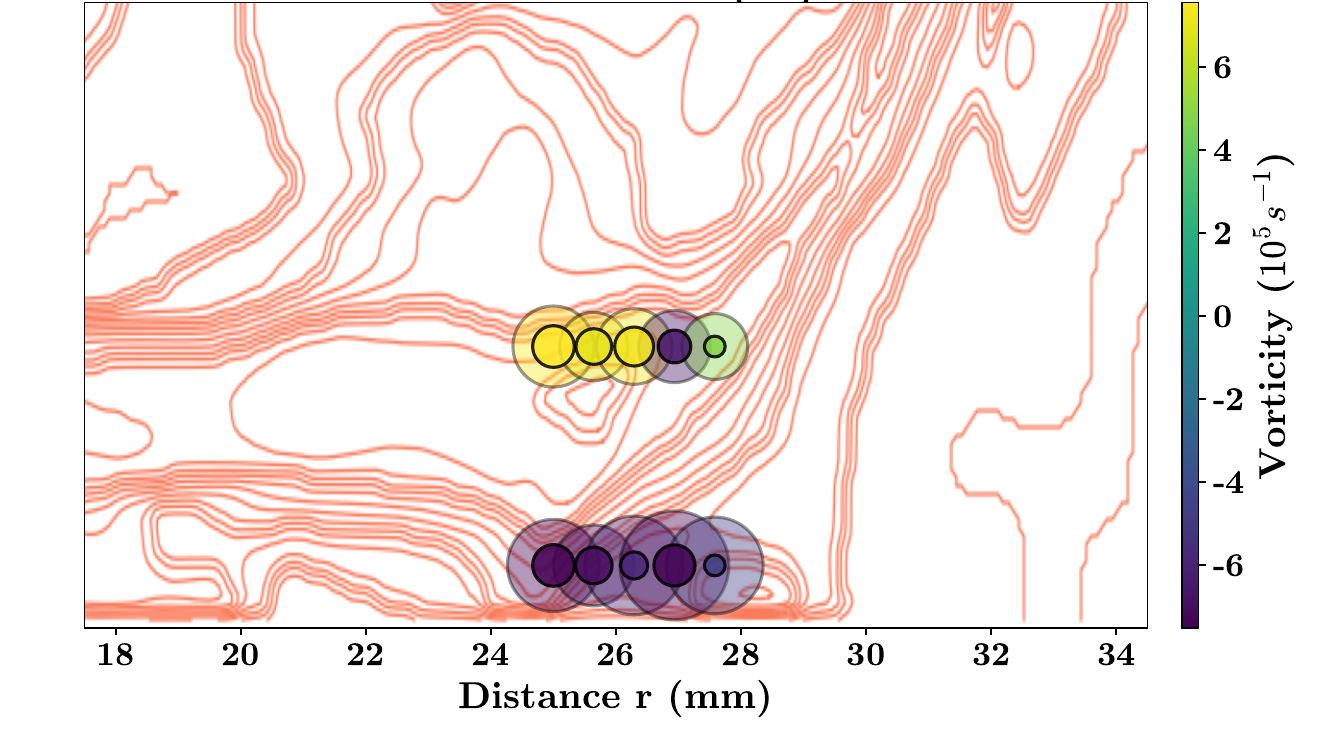}
\caption{\label{fig:Vorex}Vorticity estimates derived from axial velocity measurements for Slice 6 and Slice 10 are shown. Solid markers denote the mean vorticity obtained from 20 single-shot velocity measurements, while the halo surrounding each marker indicates $\pm2$ standard deviations of the single-shot measurements at that location. Marker area is scaled according to the relative vorticity magnitude within each slice and is overlaid on simulation velocity isoclines for context.}
\end{figure}

Based on the data in Table~\ref{tab:table4} and the scatter plot in Fig.~\ref{fig:Vorex}, a qualitative assessment of the vorticity distribution can be made. The ratio of the mean vorticity to its standard deviation, computed from $20$ single-shot CRBS measurements, is significantly larger along Slice $10$ than along Slice $6$. The measurements from the jet's core (Slice $6$), reveal a lower mean vorticity but much larger fluctuations, reflecting the highly unsteady nature of the core flow. In contrast, the shear layer (Slice 10), shows comparatively steady behavior, with the exception of one outlier visible in Fig.~\ref{fig:Vorex}. Additionally, the sign of the vorticity is mostly opposite in the two regions. These trends are consistent with observations reported in the literature~\cite{Yao_supersonic,Xiaopeng_LES}.

\section{Conclusion}
In this article, we report, for the first time to the best of our knowledge, the capability of single-shot CRBS to perform gas velocimetry in a supersonic flow regime. As previously demonstrated, this technique meets the requirements for a seedless and non-resonant optical diagnostic method capable of accurately measuring gas flow velocities. The results presented here clearly illustrate the differences between time-averaged velocity distributions and single-shot measurements in a highly underexpanded gas jet. Significant shot-to-shot variations are observed at each spatial location within the jet. Many spectra reveal regions of high gas density, characterized by pronounced Brillouin peaks—particularly in areas near shock waves, as seen in the schlieren images—indicating strongly non-Maxwellian and non-equilibrium conditions. Conversely, several spectra display well-defined CRBS line shapes, corresponding to steady supersonic flow conditions.

While the performed CFD simulations indicate that the velocities estimated from the CRBS lineshapes are comparable, these results should be interpreted with caution. Many of the spectra are acquired in the vicinity of shock waves and therefore exhibit non-Maxwellian velocity distributions. A more rigorous validation of these measurements would necessitate the use of Direct Simulation Monte Carlo (DSMC) methods, following the approach demonstrated by Wu et al.(see Ref.~\cite{wu2005parallel}), which will be pursued in future work.

This demonstration also incorporates the multipoint CRBS measurement technique, enabling simultaneous measurements of velocity and density at two distinct spatial locations~\cite{Auk_multiCRBS}. This approach facilitates the determination of velocity gradients, which can be used to estimate important turbulent flow quantities such as vorticity. Although we have demonstrated multipoint probing along a single direction, the method can be readily extended to other directions of interest by generating additional optical lattices orthogonal to the desired direction and performing single-shot CRBS with the same phase matching scheme, as recently demonstrated for subsonic flows~\cite{kumar2026simultaneous}. This extension will be explored in future works.

These features render single-shot CRBS an exceptionally versatile technique for turbulent flow velocimetry, applicable to a broad range of scenarios—from aerospace applications, such as supersonic flows, propulsion systems, and planetary entry, to the validation of models predicting non-equilibrium or non-Maxwellian behavior. When combined with density and velocity gradient estimates obtained from the single-shot CRBS lineshapes, the technique provides a powerful means to simultaneously characterize multiple flow properties within a single $250$~ns laser pulse.

\section{Acknowledgments} 
The authors would like to thank Luxembourg National Research Funds $15480342$ (FRAGOLA) and $17838565$  (ULTRAION) for supporting the research activities. The authors would also like to thank Dr. Mikhail N. Shneider for stimulating discussions.

\bibliography{sorsamp}

\end{document}